\documentclass[twocolumn,nofootinbib,floatfix,10pt,aps]{revtex4-2}
\usepackage{amsmath}
\usepackage{amssymb}
\usepackage{wasysym}
\usepackage{graphicx}
\usepackage{color,soul}
\usepackage{physics}
\usepackage{siunitx}
\usepackage{dsfont}
\usepackage{float}
\usepackage{multirow}
\usepackage{mathdots} 
\usepackage[english]{babel}
\usepackage{blindtext}
\usepackage[english,nomargin,inline,marginclue,draft]{fixme}
\pdfpageheight\paperheight
\pdfpagewidth\paperwidth
\usepackage[colorlinks,linkcolor=blue,anchorcolor=blue,citecolor=blue,urlcolor=blue]{hyperref}
            
\fxusetheme{colorsig}
\FXRegisterAuthor{cg}{acg}{CG}  
\FXRegisterAuthor{th}{ath}{\color{blue}TH}  
\FXRegisterAuthor{ib}{aib}{\color{red}IB} 
\FXRegisterAuthor{sh}{ash}{\color{cyan}SH} 
\FXRegisterAuthor{db}{adb}{\color{green}DB} 
\FXRegisterAuthor{ps}{aps}{PS}
\makeatletter
\renewcommand*\FXLayoutInline[3]{%
  {\@fxuseface{inline}\ignorespaces{\color{fx#1}[#3: #2]}}}
\makeatother

\long\def\symbolfootnote[#1]#2{\begingroup%
\def\thefootnote{\fnsymbol{footnote}}\footnotetext[#1]{#2}\endgroup}

\def\nobreakbefore{%
  \relax\ifvmode\else
    \ifhmode
      \ifdim\lastskip > 0pt\relax
        \unskip\nobreakspace
      \else 
        \nobreakspace
      \fi
    \fi
  \fi
}
\let\oldcite\cite
\renewcommand\cite{\nobreakbefore\oldcite}

\begin{document}

\title{Fingerprint Recognition of Partial Discharge Signals in Deep Learning Enhanced Rydberg Atomic Sensors}

\author{Yi-Ming Yin$^{1,2}$}
\author{Qi-Feng Wang$^{1,2}$}
\author{Yu Ma$^{1,2}$}
\author{Tian-Yu Han$^{1,2}$}
\author{Jia-Dou Nan$^{1,2}$}
\author{Zheng-Yuan Zhang$^{1,2}$}
\author{Han-Chao Chen$^{1,2}$}
\author{Xin Liu$^{1,2}$}
\author{Shi-Yao Shao$^{1,2}$}
\author{Jun Zhang$^{1,2}$}
\author{Qing Li$^{1,2}$}
\author{Ya-Jun Wang$^{1,2}$}
\author{Dong-Yang Zhu$^{1,2}$}
\author{Qiao-Qiao Fang$^{1,2}$}
\author{Chao Yu$^{1,2}$}
\author{Bang Liu$^{1,2}$}
\author{Li-Hua Zhang$^{1,2}$}
\author{Dong-Sheng Ding$^{1,2}$}
\email{dds@ustc.edu.cn}
\author{Bao-Sen Shi$^{1,2}$}

\affiliation{$^1$Key Laboratory of Quantum Information, University of Science and Technology of China, Hefei, Anhui 230026, China.}
\affiliation{$^2$Synergetic Innovation Center of Quantum Information and Quantum Physics, University of Science and Technology of China, Hefei, Anhui 230026, China.}
\affiliation{}
\date{\today}

\begin{abstract}

Partial discharge originates from microscopic insulation imperfections in high-voltage apparatus and is widely considered a critical marker of incipient deterioration. Conventional partial discharge detection methods are typically constrained by limited bandwidth and often rely on predefined feature extraction, which impedes reliable recognition of broadband transient signals. In this work, we employ a Rydberg atomic sensor to directly capture time-domain responses of partial discharge emissions and construct distinctive spectral fingerprints for different types. A 1D ResNet deep learning model is then applied to recognize these fingerprints from time-domain signals without manual feature engineering. Under increased source–antenna distances, where spectral features are significantly attenuated, the model attains a recognition accuracy of approximately 94$\%$ across four partial discharge categories, demonstrating robustness to attenuation and noise. We further validate the approach in a simulated early-warning scenario, where partial discharge signals mixed with noise are analyzed and the model successfully generates predictive alarms. These results underscore the potential of integrating Rydberg-based broadband sensing with data-driven analysis for non-invasive, high-sensitivity diagnostics of electrical insulation systems.

\end{abstract}

\maketitle

\section{Introduction}

Partial discharge is a localized electrical breakdown in the insulation of high-voltage equipment and is widely regarded as an early indicator of insulation degradation\cite{Hussain2023, Wong2015}. It produces transient electromagnetic pulses with sub-nanosecond rise times and broadband spectra spanning from the megahertz to gigahertz range. Given that different partial discharge types exhibit distinct spectral signatures\cite{Robles2018}, broadband measurement is crucial for capturing their discriminatory features and enabling reliable recognition.

Conventional partial discharge detection techniques, particularly Ultra-High Frequency (UHF) sensors and pulse current methods, have achieved high recognition rates when combined with deep learning approaches\cite{Dalila2025, Evagorou2010, Wang2022}. However, these hardware approaches are often constrained by limited bandwidths determined by sensors geometry and require complex calibration chains. In contrast, Rydberg-atom-based electric field sensors have recently emerged as a promising platform for broadband diagnostics. By leveraging electromagnetically induced transparency (EIT)\cite{Liao2020, Fleischhauer2005, AbiSalloum2010, Foot1998}, the ac Stark shift\cite{ Foot1998,AutlerTownes1955,BohlouliZanjani2007}, and Autler–Townes splitting\cite{ AbiSalloum2010, Foot1998}, these sensors derive distinct physical advantages from fundamental physical constants, specifically intrinsic SI-traceability and self-calibration capabilities\cite{Holloway2014} alongside an ultra-wide dynamic frequency response range extending from MHz to THz\cite{Bason2010,LiuSensitive,Meyer2020,Meyer2021,Wang2025ArcRF,ZHANG2024100089}. Furthermore, their nonmetallic structure and isotropic response allow them to couple directly to fields without metallic probes\cite{Holloway2017,Sedlacek2012}, enabling the atomic response to intrinsically map incident fields into spectral fingerprints.

To recognize these spectral fingerprints, we adopt a 1D ResNet-based deep learning model capable of accurately analyzing the time-domain signals extracted from the Rydberg atom sensors without manual feature engineering\cite{he_resnet,Ismail_Fawaz2019,ismail2020inceptiontime,Pedregosa_sklearn_2011,rs16213986,Mao2021RadarSM}. The model updates its weights through backpropagation, autonomously learning relevant representations from large datasets without requiring prior knowledge of the underlying physics or the experimental system. Leveraging these advantages, deep neural networks have been successfully applied in fields demanding subtle pattern recognition from noisy measurements, including multifrequency microwave recognition\cite{Liu2022}, far-field subwavelength acoustic imaging\cite{Orazbayev2020PhysRevX}, time-dependent atomic magnetometry\cite{Khanahmadi2021PhysRevA}, optical vortex recognition\cite{ Giordani2020PhysRevLett, Liu2019PhysRevLett}, and automatic quantum control\cite{Wigley2016SciRep,Mukherjee2020PhysRevLett,Tranter2018NatCommun,Mills2020NatMachIntell,Wang2020PhysRevLett,Bukov2018PhysRevX}.
    
In this work, we present a recognition scheme that combines Rydberg-atom sensing with a 1D ResNet model. The atomic system encodes partial discharge signatures as spectral fingerprints, which weaken with increasing source–sensor distance. To ensure reliable identification, we train the model on experimental fingerprint datasets, achieving robust recognition even under low-signal conditions. Tests with signals mimicking realistic discharge scenarios further demonstrate its capability for early warning, coupling the promise of quantum sensors with deep learning for broadband, non-invasive diagnostics.

\begin{figure*}[htb]
    \centering\includegraphics[width=2\columnwidth]{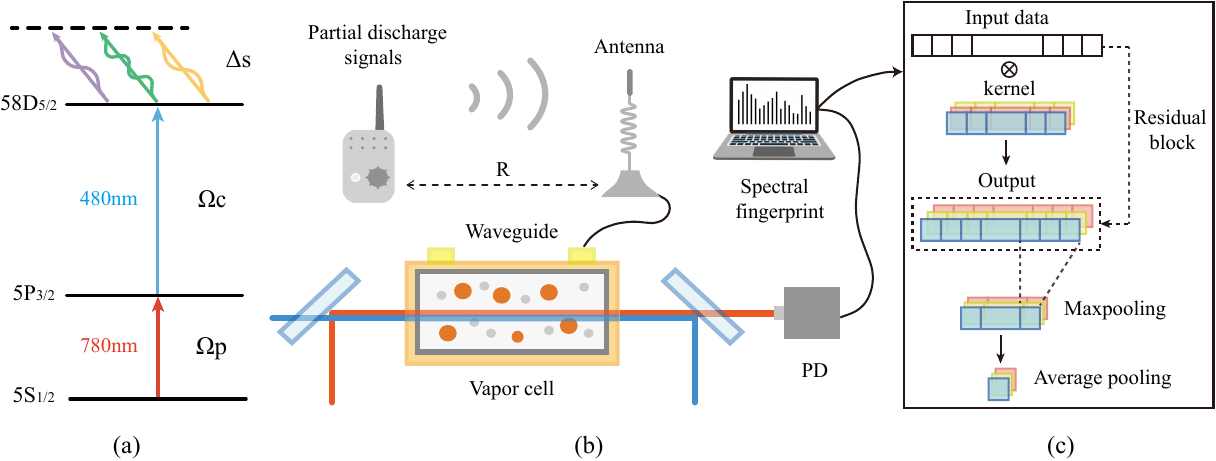}
    \caption{\textbf{Illustration of the setup and physical diagram.} (a) Overview of experimental energy diagram. The probe (780nm) and coupling (480nm) laser couple the atomic states of the ground state $5\rm{S}_{1/2}$, the intermediate state $5\rm{P}_{3/2}$ and the Rydberg state $58\rm{D}_{5/2}$, forming a ladder-type electromagnetically induced transparency (EIT) scheme. In the presence of partial discharge fields, the Rydberg level $58\rm{D}_{5/2}$ exhibits an energy shift $\Delta s$ due to the ac Stark effect. (b) Experimental setup. Partial discharge signals are received by an antenna at different distances R, coupled into a waveguide, and radiated directly to the Rb vapor cell. The EIT transmission of the probe beam is detected by a photodetector (PD) to obtain spectral fingerprints. (c) Schematics of the residual neural network. The network comprises a one-dimensional convolution layer, residual blocks, a maxpooling layer and a global average pooling layer. For further details about these layers, see the ``Method’’ section.}
    \label{fig1}
\end{figure*}

\section{Physical model}

The experimental setup and the experimental energy diagram are illustrated in Figures.~\ref{fig1}(a-b). An ultra-wideband (UWB) antenna (950 MHz–7 GHz) functions as a front-end collector to enhance the coupling efficiency of partial discharge radiation into the waveguide (300 MHz–24 GHz) housing the Rb vapor cell. At the atomic level, $^{85}\text{Rb}$ atoms are excited from the ground state $5\text{S}_{1/2}$ to the Rydberg state $58\text{D}_{5/2}$ via a two-photon EIT process. Consequently, the coherent interaction between the Rydberg atoms in the three-level ladder system $\{|g\rangle, |e\rangle, |r\rangle\}$ and the oscillating electric fields delivered by the UWB antenna can be described by the total Hamiltonian
    \begin{equation}
        H = \frac{\hbar}{2} \Big( \Omega_p |g\rangle\langle e| + \Omega_c |e\rangle\langle r| + \text{H.c.} \Big) + \Delta_s |r\rangle\langle r|, 
    \end{equation}
where $\Omega_p$ and $\Omega_c$ are the probe and coupling Rabi frequencies, H.c. denotes Hermitian conjugate, and $\Delta_s$ denotes the ac Stark shift induced by the partial discharge field. The Stark perturbation can be approximated as
    \begin{equation}
        \Delta_s \approx \alpha(\omega) |E_t|^2,
    \end{equation}
where $E_t$ is the partial discharge field amplitude and $\alpha(\omega)$ is the frequency-dependent polarizability[4]. Since the polarizability scales as $\alpha(\omega)\propto n^{7}$, the choice of principal quantum number involves a trade-off between sensitivity and environmental susceptibility. In this work, the $58\text{D}_{5/2}$ state is selected to balance the high sensitivity required for weak signal detection with the spectral stability needed to prevent state mixing. This Stark shift modifies both the central frequency and linewidth of the EIT resonance, effectively transforming the probe susceptibility into $\chi(\Delta_p, E_{t})$, which encodes the spectral fingerprints of different partial discharge types.
    
To decode these fingerprints, we employ a residual neural network as illustrated in Fig.~\ref{fig1}(c), which realizes a nonlinear mapping
    \begin{equation}
    	f_\theta: \ \chi(\Delta_p, E_{t}) \mapsto y,
    \end{equation}  
where $\chi(\Delta_p, E_{t})$ denotes the measured fingerprint modified by the Stark effect, $y \in \{0,1,2,3\}$ indicates the partial discharge category, and $\theta$ represents the trainable parameters. 

Physically, the shallow convolutional layers of the network are sensitive to local oscillatory features arising from sharp Stark-induced modulations of the EIT resonance, while deeper layers capture global structures such as peak distributions and linewidth variations. The convolution operation in each layer can be formally expressed as
    \begin{equation}
        h_j^{(l+1)}(t) = \sigma \Bigg( \sum_i \int K_{ij}^{(l)}(\tau)\, \chi_i(t-\tau)\, d\tau + b_j^{(l)} \Bigg),
    \end{equation}
where $l$ denotes the layer index, $i$ and $j$ denote the input and output channels, $K_{ij}^{(l)}(\tau)$ represents the convolution kernel capturing local features, $b_j^{(l)}$ is the bias, and $\sigma$ is the activation function. This formulation directly links the atomic response to the hierarchical feature extraction in the network, providing a physical interpretation of how local Stark-induced oscillations and global spectral patterns are encoded into the learned representations.

\begin{figure*}[htb]
    \centering\includegraphics[width=2\columnwidth]{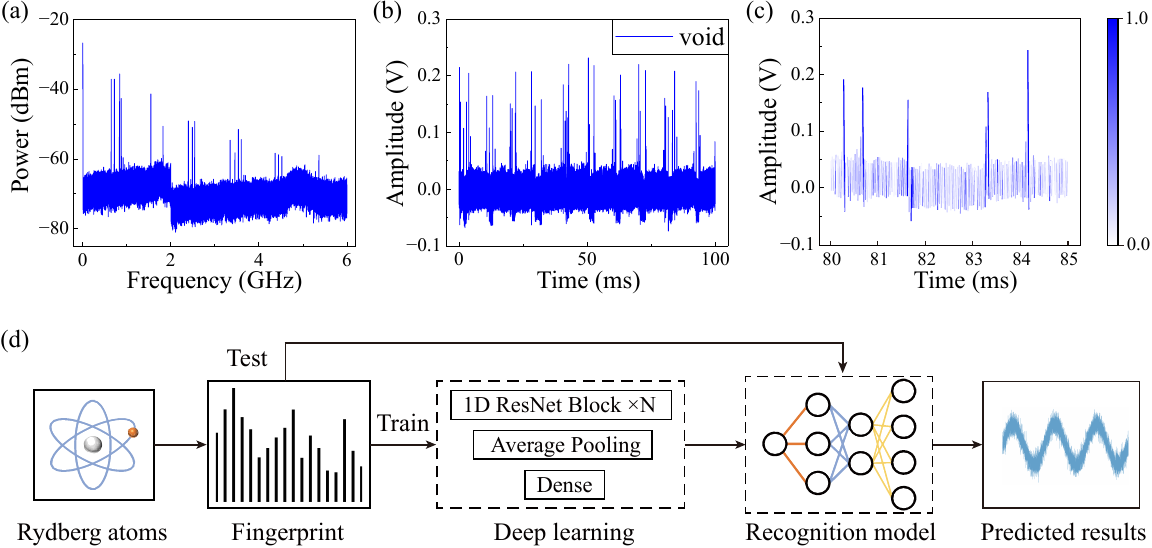}
    \caption{\textbf{Spectral fingerprint construction and recognition process.} (a) Measured frequency spectrum of the void-type partial discharge signal spanning 0-6GHz. The data was obtained by directly connecting the discharge source to a spectrum analyzer to characterize the intrinsic frequency content of the source. (b) Time-domain waveform extracted from the Rydberg atom sensors when the partial discharge source is 1 cm from the antenna. (c) Saliency map visualization highlighting the model’s attention across different temporal regions of the signal. (d) Flowchart of fingerprint recognition. Partial discharge signals are transmitted via an antenna to the Rydberg atom system, and encoded as spectral fingerprints, which are then divided into training and test sets. The training set is used to train a residual network, generating a recognition model, which is then evaluated on the test set to assess recognition performance. For further details about the training and testing process, see the ``Method’’ section.}
    \label{fig2}
\end{figure*}

\section{Results}

\subsection{The construction and recognition of fingerprints}
The spectral characteristics of each type of partial discharge signal are inherently distinct, resulting in time-domain waveforms extracted from the Rydberg atom sensor that differ accordingly. Figures.~\ref{fig2}(a–b) present the frequency range and the corresponding time-domain waveform of a void-type partial discharge signal, respectively. To investigate how the deep learning model constructs spectral fingerprints from these time-domain waveforms, saliency maps were introduced to visualize the model’s attention across different temporal regions, as shown in Fig.~\ref{fig2}(c). In the saliency maps, the color intensity represents the degree of model attention at each time point: moments exhibiting significant amplitude deviations or abrupt changes typically correspond to deeper colors. This indicates that the model primarily focuses on local extrema or transient features within the waveform. These distinctive transient features carry the unique characteristics of each partial discharge type, enabling the model to extract and integrate them into spectral fingerprints that distinguish different partial discharge signal classes.
   
Figure.~\ref{fig2}(d) illustrates the overall spectral fingerprint recognition process. In this procedure, partial discharge signals are first captured via an antenna and converted into time-domain waveforms by the Rydberg atom sensing system. The deep learning model then processes these waveforms to extract relevant features and encode them into spectral fingerprints. The resulting fingerprints are partitioned into training and test sets: the training set is used to train a residual network, while the test set is employed to evaluate the recognition performance. This workflow allows for the effective classification of partial discharge signal types and ensures that the spectral fingerprints faithfully represent the distinguishing characteristics of each signal class. 

\begin{figure*}[htb]
    \centering\includegraphics[width=2\columnwidth]{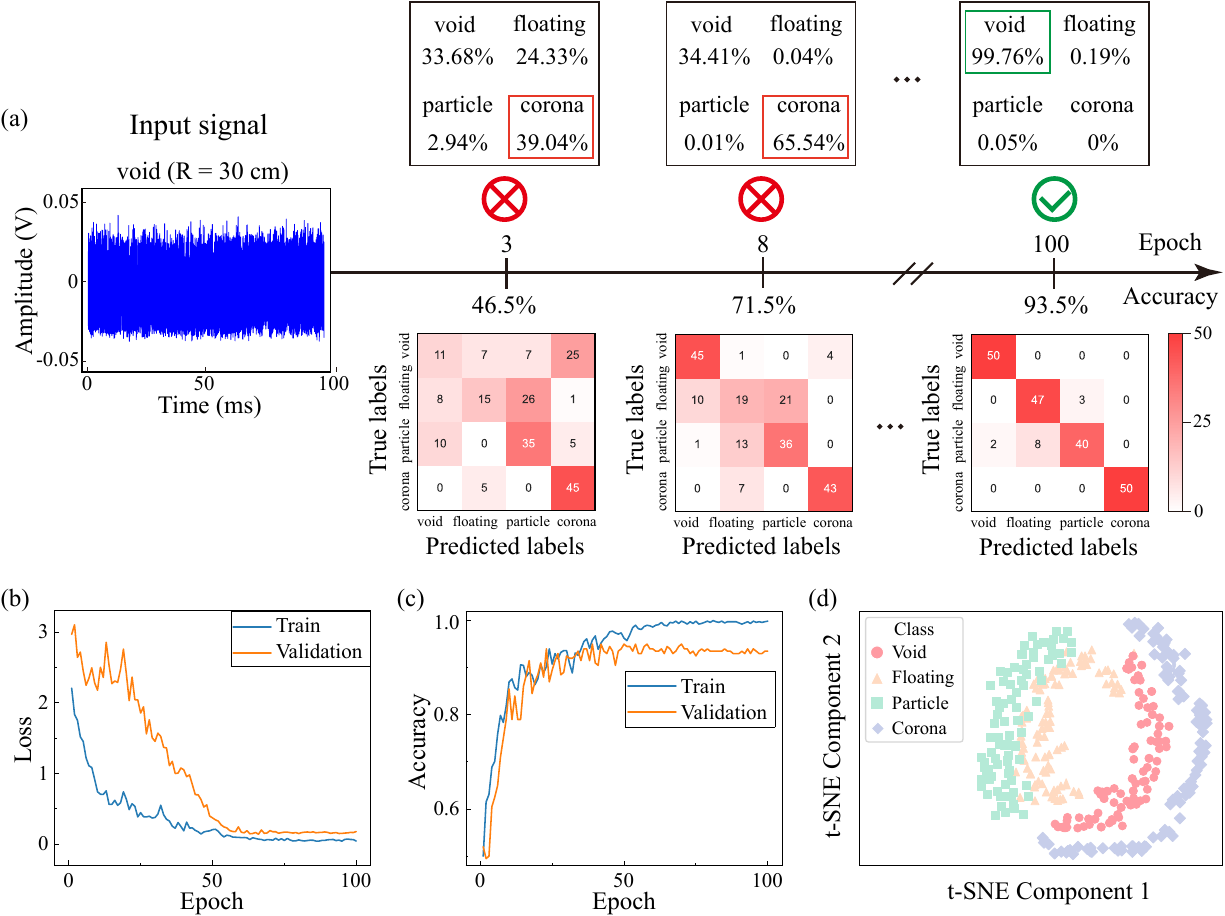}
    \caption{\textbf{Recognition performance of the deep learning model across training epochs.} (a) Confusion matrices of signals measured at 30 cm after 3, 8, and 100 training epochs, with overall recognition accuracies of 46.5$\%$, 71.5$\%$, and 93.5$\%$, respectively. (b-c) Evolution of loss and accuracy curves with epoch for the training set (blue) and the validation set (orange). (d) t-SNE visualization of the extracted feature representations.}
    \label{fig3}
\end{figure*}

\subsection{Deep learning}
As the distance between the antenna and the partial discharge source increases, the signal undergoes rapid attenuation accompanied by rising noise, which causes the signal-to-noise ratio (SNR) to drop significantly. The SNR is defined as
    \begin{align}
        \mathrm{SNR} = 20 \log_{10}\left(\frac{A_{\text{signal}}}{A_{\text{noise}}}\right),
    \end{align}
When the distance increases from 1 cm to 30 cm, the SNR decreases from 16 dB to 2 dB. Under such conditions, the distinctive features of the signals become too weak to be visually distinguished, necessitating the use of deep learning models to extract their latent representations for reliable recognition.
    
To evaluate the recognition performance under low-SNR conditions, we collected 1000 signal samples at a distance of 30 cm, with 250 signals from each partial discharge type. A standard 80/20 split was applied, allocating 80$\%$ of the data for training and 20$\%$ for testing. The recognition results at different training stages are summarized in Fig.~\ref{fig3}(a). At an early stage of training (epoch 3), the network misclassifies the signal as corona, yielding a low overall accuracy of 46.5$\%$ as shown by the matrix. By epoch 8, the prediction is still incorrect but the overall accuracy improves to 71.5$\%$. After sufficient training (epoch 100), the model correctly identifies the signal as void, with the classification accuracy increasing to 93.5$\%$. These results highlight how the residual network gradually extracts discriminative spectral features from fingerprints, eventually achieving robust recognition despite the low SNR at longer distances.

\begin{figure*}[htb]
    \centering\includegraphics[width=1.5\columnwidth]{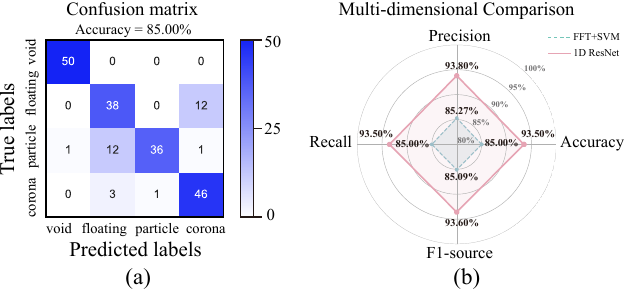}
    \caption{\textbf{Performance comparison between the classical baseline and the deep learning method.} (a) Confusion matrix of the FFT+SVM baseline, showing significant misclassifications between floating and particle discharges. (b) Multi-dimensional comparison radar chart of FFT+SVM (dashed line) versus 1D ResNet (solid line) across four evaluation metrics, demonstrating the comprehensive superiority of the deep learning approach.}
    \label{fig4}
\end{figure*}
    
The training dynamics are illustrated in Fig.~\ref{fig3}(b–c). Due to the extremely low SNR, the model requires nearly 50 epochs before reaching convergence, and the learning curves exhibit pronounced oscillations before stabilization. Eventually, the training accuracy approaches nearly 100$\%$, while the validation accuracy consistently stabilizes at about 94$\%$, suggesting effective learning but also limited generalization. The t-SNE visualization in Fig.~\ref{fig3}(d) further reflects this trend, where the four signal classes are still separable but the clusters appear less compact and partially overlapping, suggesting that noise contamination reduces class separability in the latent space. Collectively, these results indicate a mild degree of overfitting: the model fits the training set well but exhibits limited generalization to unseen data. The primary factor is not insufficient training samples, as both datasets contain 1000 signals, but rather the reduced signal-to-noise ratio at a longer distance, which diminishes the distinctiveness of discriminative features. Consequently, the learning task becomes intrinsically more difficult when critical features are embedded in noise, and the achievable performance is ultimately constrained by the physical quality of the input rather than by data volume or network capacity.

To rigorously validate that the network’s performance is driven by learned physical features rather than overfitting, we benchmarked our 1D ResNet against a classical FFT+SVM baseline. As shown in Fig.~\ref{fig4}(a), the SVM baseline achieves a robust accuracy of 85$\%$, confirming that the Rydberg dataset contains physically informative fingerprints even under low-SNR conditions. However, the confusion matrix reveals that the classical approach struggles to distinguish between "floating" and "particle" discharges, which account for the majority of misclassifications.

Crucially, our deep learning model outperforms this baseline by a significant margin of 8.5$\%$, achieving a nearly uniform improvement across all categories. This comprehensive superiority is visually summarized in the multi-dimensional comparison in Fig.~\ref{fig4}(b), where the 1D ResNet profile completely encloses the baseline across all metrics (Accuracy, Precision, Recall, and F1-score). This comparison clarifies that the challenges in cluster compactness observed earlier are due to the intrinsic physical complexity of the signals rather than model overfitting. The results demonstrate that the residual network effectively captures complex, non-linear dependencies within the Rydberg fingerprints that are inaccessible to standard spectral analysis, thereby justifying the necessity of the proposed deep learning architecture.

\subsection{Early warning based on spectral fingerprints}
\begin{figure*}[htb]
    \centering\includegraphics[width=2\columnwidth]{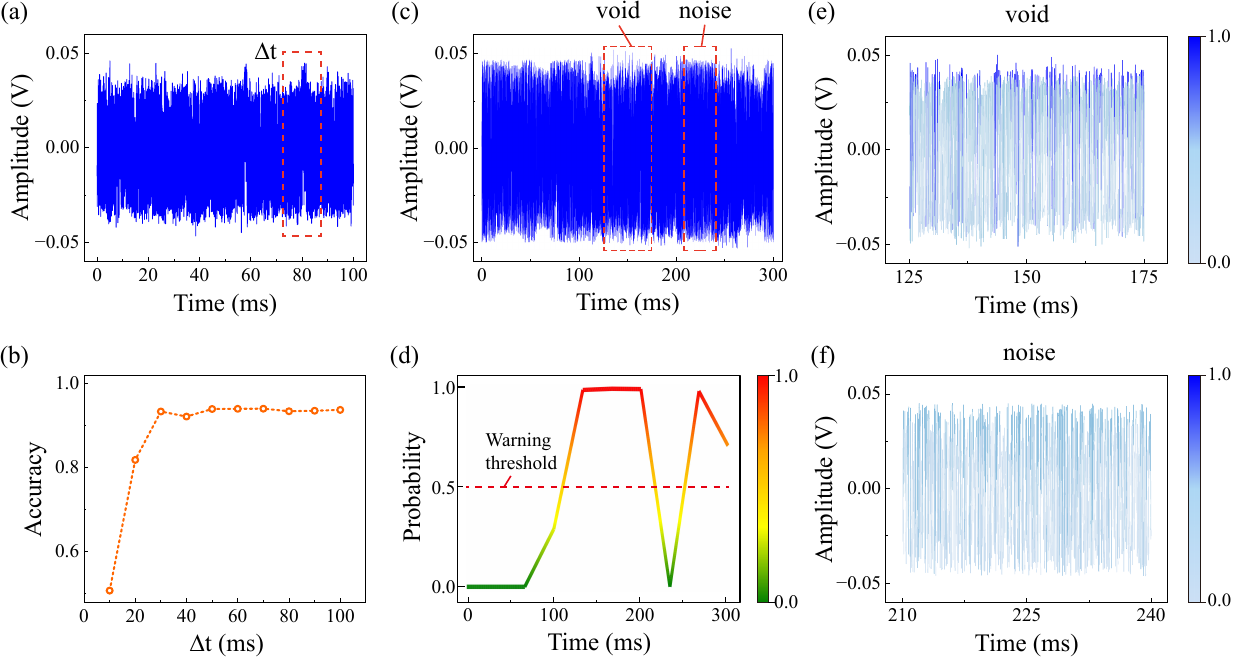}
    \caption{\textbf{Prediction performance and early-warning of partial discharge signals.} (a) Time-domain waveform of the void-type partial discharge signal measured at 30 cm. (b) Prediction accuracy versus time window length ($\Delta$t). (c) A 300 ms signal composed of void signal and noise. (d) Probability versus time for early-warning detection of the signal shown in (c), with 0.5 set as the alarm threshold. (e-f) Saliency map of the void signal and noise.}
    \label{fig5}
\end{figure*}

To assess the feasibility of applying spectral-fingerprint recognition to practical monitoring scenarios, we carried out a simulation of early warning. Figure.~\ref{fig5}(a) presents a typical temporal waveform of a void-type partial discharge measured at 30 cm. A variable time window $\Delta$t was extracted and used as the model input, with the prediction accuracy shown in Fig.~\ref{fig5}(b). The results reveal a clear dependence on the observation time: while shorter windows provide limited discriminatory information, the accuracy rapidly improves with increasing $\Delta$t and exceeds 90$\%$ once $\Delta$t $\geq$ 30 ms. This behavior indicates that the spectral fingerprints, even when partially sampled, contain sufficient distinctive features for reliable recognition, highlighting the robustness of the Rydberg-based sensing scheme.

To emulate a realistic early-warning scenario, we further constructed a composite signal consisting of void-type discharges embedded in stochastic noise as shown in Fig.~\ref{fig5}(c). The model predictions, visualized in Fig.~\ref{fig5}(d), are mapped to a probability scale ranging from 0 (green) to 1 (red), with 0.5 defined as the alarm threshold. The model’s output probabilities are highly polarized, with discharge events consistently scoring near 1 and noise near 0. The distinct separation validates 0.5 as an effective cutoff for generating predictive alarms in the current setup. Furthermore, this threshold is adjustable in practical applications, enabling a flexible trade-off between precision and recall to suit different operational environments.
    
To validate the reliability of the warning mechanism, representative segments corresponding to discharge and non-discharge events were further analyzed in Fig.~\ref{fig5}(e–f). The accompanying attention maps reveal the temporal regions most influential for the model’s decision-making. Notably, the highlighted regions coincide with the characteristic transients of partial discharge pulses. This observation aligns with the physical mechanism of our sensing approach, wherein the atomic AC Stark shift physically encodes spectral features into the time-domain waveforms extracted from Rydberg sensor, thereby constituting the “spectral fingerprints” analyzed in this work. The correspondence between the attention weights and these physical features provides strong evidence that the model does not rely on spurious correlations but instead correctly identifies the underlying spectral-temporal fingerprints.

\section{Discussion}

In this work, we establish a recognition framework that combines Rydberg-atom sensing with a 1D ResNet model to analyze partial discharge signals. By encoding discharge events as spectral fingerprints, the atomic system provides broadband, non-invasive access to transient electromagnetic fields. Using this framework, we achieved classification of four representative partial discharge types with an accuracy of about 94$\%$. The model also maintains high performance across different temporal windows, exceeding 90$\%$ once the observation time reaches 30 ms, and remains effective under noise-contaminated conditions. Furthermore, the early warning simulation demonstrates that discharge events can be reliably distinguished from noise with a probability-based threshold, while attention maps confirm that the model's decisions are rooted in physically meaningful transients.
    
In summary, we demonstrate that Rydberg-atom-based spectral fingerprinting, combined with deep learning, enables robust recognition and early warning of partial discharge signals, highlighting the potential of quantum-enhanced sensing as a broadband, non-invasive diagnostic tool. To fully exploit this intrinsic ultra-wideband nature and minimize invasiveness, future work aims to replace the current metallic antenna coupling with an antenna-free, all-dielectric architecture. Furthermore, to facilitate practical deployment in varying environments\cite{Zhang2021ResNetRF,Soltanieh2020Review,Shen2024FedRFFI,Zhang2023LengthRobustRFFI}, such as those with variable loads or unknown discharge behaviors\cite{Hu_SE_2018}, the system can adopt incremental or transfer learning strategies\cite{ijcai2018p369}. These approaches allow the pre-trained backbone to be efficiently fine-tuned using a small number of new samples, thereby enabling rapid adaptability to complex, realistic scenarios without the need for extensive retraining.

\section{Method}
\subsection{Deep learning layers}
Our deep learning model consists of a 1D CNN layer, three ResidualBlock layers and a dense layer\cite{Srivastava_Dropout_2014,Huang_ResNet_Dropout_2016,Wan_DropConnect_2013}. The mathematical sketches for the key components are as follows.

Given an input waveform $\textbf{x} \in \mathbb{R}^{1 \times L}$, the model first applies a convolutional stem composed of four operations: 1D convolution, batch normalization\cite{Ioffe_BN_2015}, ReLU activation, and max pooling. The convolutional layer extracts local features using
    \begin{align}
        z^{c}(n) = (\textbf{x}\otimes W^{c})(n) = \sum_{m=0}^{k-1}\textbf{x}_{(n+m)}\cdot W_{m}^{c},
    \end{align}
where $\textbf{x}$ represents the input signal, $W^{c}$ is the convolution kernel for the output channel \emph{c}, \emph{m} is the kernel index, and \emph{n} is the output temporal index.  

Batch normalization and ReLU are then applied in turn, i.e. $\hat{z}=(z-\mu_{B})/\sqrt{\sigma_B^2+\epsilon}$, $y=\gamma \hat{z}+\beta$, and $f(x)=\max(0,x)$. Finally, max pooling with kernel size 3 is performed as 
    \begin{align}
        y(m) = \max_{i\in [m,m+k-1]}x(i),
    \end{align}
which reduces the temporal resolution while preserving dominant features.

The core of the network consists of three stacked residual blocks. Each block includes two convolutional layers (kernel size 3), each followed by batch normalization and ReLU. To mitigate degradation in deep networks, residual connections add the block's input directly to its output:
    \begin{align}
        \textbf{y}={\rm ReLU}(z_{2}+\textbf{\textit{S}}(\textbf{x})),
    \end{align}
where $\textbf{\textit{S}}(\textbf{x})=\textbf{x}$ for matching dimensions, or $\textbf{\textit{S}}(\textbf{x})={\rm BN(Conv1D_{1\times1}(\textbf{x}))}$ otherwise.

The output of the final residual block, $\textbf{Z}\in\mathbb{R}^{C\times T}$, is processed by adaptive average pooling to produce a fixed-size vector $v_{c}=\frac{1}{T}\sum_{t=1}^{T}Z_{c,t}$, yielding $\textbf{v}\in\mathbb{R}^{C}$ with $C=256$ in our implementation. The final prediction is obtained via a fully connected layer:
    \begin{align}
        \textbf{y}=\textbf{W}\cdot\textbf{v}+\textbf{b},
    \end{align}
followed by the softmax function
    \begin{align}
        \hat{y}_{i}=\frac{e^{y_{i}}}{\sum_{j=1}^{K}e^{y_{j}}},
    \end{align}
which gives class probabilities.

The model is trained by minimizing the categorical cross-entropy loss
    \begin{align}
        \mathcal{L}_{CE}=-\sum_{i=1}^{K}y_{i}\log(\hat{y}_{i}),
    \end{align}
where $y_i \in \{0,1\}$ is the one-hot ground truth and $\hat{y}_i$ is the predicted probability for class $i$. Parameters are updated with Adam optimizer.

The network is implemented using the PyTorch 2.6.0 framework on Python 3.12.6\cite{ Ansel_PyTorch2_2024}. All weights are initialized with the PyTorch default. The hyper-parameters of the deep learning model (including the convolution kernel size, the number of channels, and the learning rate) are tuned empirically based on validation performance.

\begin{figure}[htb]
    \centering\includegraphics[width=0.8\columnwidth]{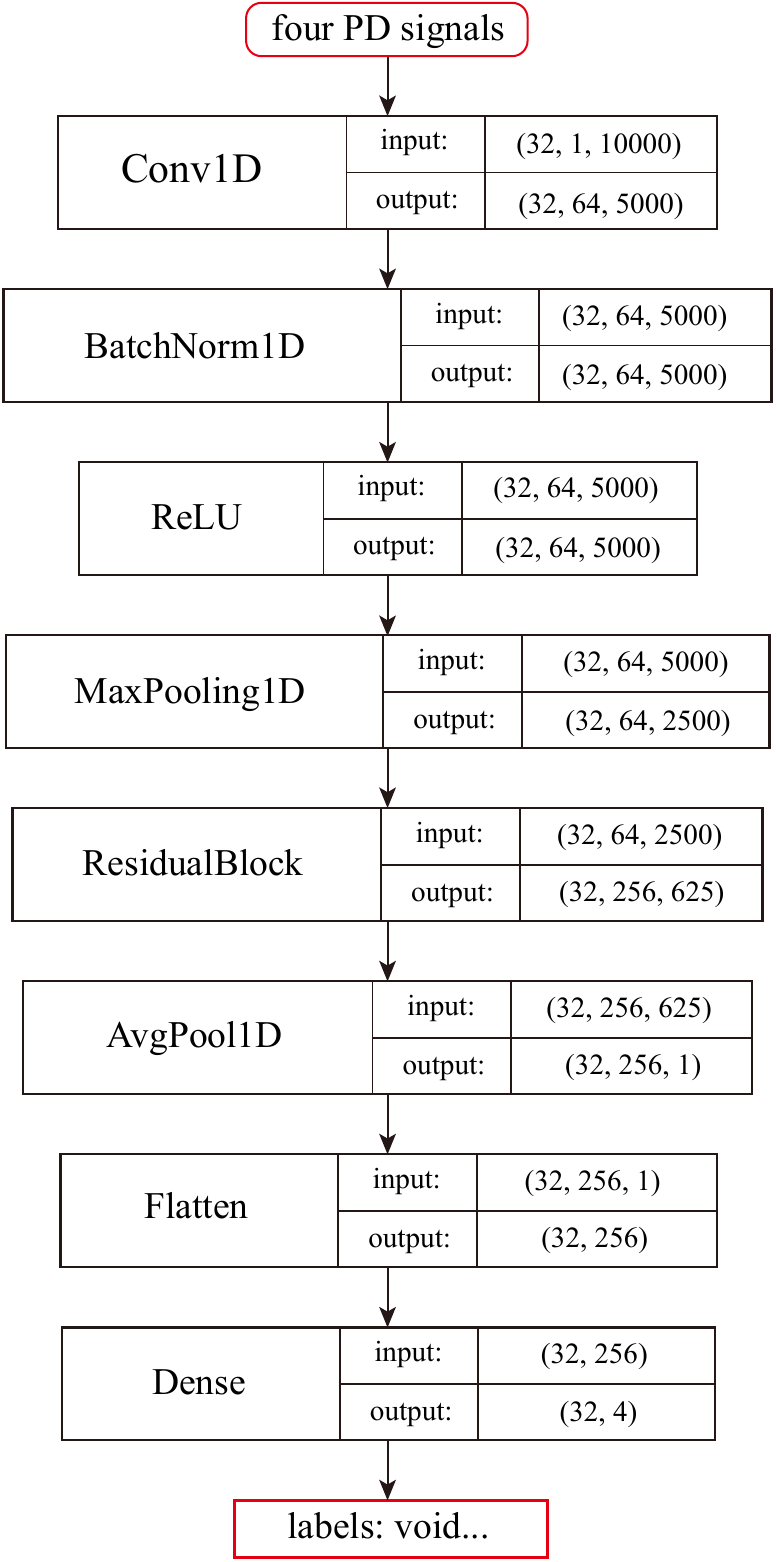}
    \caption{\textbf{Structure of our deep learning model and size of the data.} Each batch contains 32 samples with input shape (1, 10000). The network consists of a convolutional stem, batch normalization, ReLU, pooling layers, a residual block, and a fully connected output layer. Feature dimensions are progressively reduced to a 4-class output.}
    \label{fig6}
\end{figure}

\subsection{Deep learning pipeline}
To ensure consistency across samples, all raw waveform signals are resampled to a fixed length of 10,000 points. The data are then standardized using a \emph{StandardScaler}, which is fitted on the training set and saved for inference to preserve distribution alignment. The labels are encoded as one-hot vectors with four elements, corresponding to the four categories of partial discharge signals.
    
The deep learning moedel structure is shown in Fig.~\ref{fig6}. The input tensor is formatted as(batch size, number of channels, signal length). In our implementation, the batch size is set to 32, the number of channels is 1, and the signal length is 10,000, resulting in an input shape of (32, 1, 10000). The deep learning model is trained for 50 epochs using the Adam optimizer with an initial learning rate of $10^{-4}$ and a weight decay of $10^{-5}$. The objective function is categorical cross-entropy loss.

To improve generalization and accelerate convergence, a learning rate scheduler \emph{ReduceLROnPlateau} is employed. The learning rate $\eta$ is reduced by a factor $\gamma$ whenever the validation loss fails to decrease for $p$ consecutive epochs:
    \begin{align}
        \eta\leftarrow\gamma\cdot\eta\  \mathrm{if\ no\ improvement\ for\ \emph{p} \ epochs}.
    \end{align}
In our experiments, we set $\gamma = 0.1$ and $p = 3$. With this configuration, on our experimental platform, the model typically converges within 100 epochs, requiring approximately 10 minutes. This low time cost facilitates rapid model iteration and potential on-site updates.

The dataset is randomly split into training and validation sets with an 80:20 ratio using stratified sampling to preserve class distribution. The same data split is used throughout all training epochs to ensure consistency. In each epoch, the model is trained on the training set and evaluated on the validation set. After training, the model parameters and fitted scaler are saved for downstream testing and deployment.

\subsection{Electric field modulation of Rydberg-EIT coherence}
The dynamics of the three-level ladder system $\{|g\rangle, |e\rangle, |r\rangle\}$ under the action of probe and coupling lasers, together with the external partial discharge field, are governed by the Lindblad master equation
    \begin{equation}
        \dot{\rho} = -\frac{i}{\hbar}[H,\rho] + \mathcal{L}[\rho],
    \end{equation}
where $\rho$ is the atomic density matrix and $\mathcal{L}[\rho]$ represents relaxation and dephasing processes.
    
In the rotating-wave approximation, the Hamiltonian of the system can be written in matrix form as
    \begin{equation}
        H = \frac{\hbar}{2}
    \begin{pmatrix}
        0 & \Omega_p & 0 \\
        \Omega_p & -2\Delta_p & \Omega_c \\
        0 & \Omega_c & -2(\Delta_p+\Delta_c+\Delta_s)
    \end{pmatrix},
    \end{equation}
where $\Omega_p$ and $\Omega_c$ are the probe and coupling Rabi frequencies, $\Delta_p$ and $\Delta_c$ denote the probe and coupling detunings, and $\Delta_s$ is the ac Stark shift induced by the PD field.The Stark perturbation is approximated as
    \begin{equation}
        \Delta_s(t) \approx \alpha(\omega)|E_t(t)|^2,
    \end{equation}
with $\alpha(\omega)$ the frequency-dependent polarizability and $E_t(t)$ the amplitude of the PD field.
    
Solving the master equation in the weak-probe limit ($\Omega_p \ll \Omega_c,\gamma$), the steady-state coherence between $|g\rangle$ and $|e\rangle$ is obtained as
    \begin{equation}
        \rho_{ge}(\Delta_p,t) =
        \frac{i\Omega_p/2}{\gamma_{eg}-i\Delta_p\dfrac{|\Omega_c|^2/4}{\gamma_{rg}-i(\Delta_p+\Delta_c+\Delta_s(t))}},
    \end{equation}
    
The linear susceptibility of the medium is then given by
    \begin{equation}
        \chi(\Delta_p,t) = \frac{N |d_{ge}|^2}{\varepsilon_0 \hbar}\rho_{ge}(\Delta_p,t),
    \end{equation}
where $N$ is the atomic density and $d_{ge}$ is the dipole matrix element of the probe transition.
    
Finally, the probe transmission through the vapor cell of length $L$ is expressed as
    \begin{equation}
        T(\Delta_p,t) = \exp\left[-kL \mathrm{Im} \chi(\Delta_p,t)\right],
    \end{equation}
with $k$ the probe wave number.

\bibliography{ref}  

\end{document}